\documentclass[a4paper]{article}

\usepackage{amsmath,amsfonts,amssymb,amsthm,graphicx,cite}

\begin{document}
\begin{center}

{\Large \bf Non-commutative geometry and exactly solvable
systems}\footnote{Contribution to the "International Conference on
Noncommutative Geometry and Physics", April 2007, Orsay (France).}
\vspace{1 cm}

{\large Edwin Langmann}\\

\vspace{0.3 cm} 

{\em Theoretical Physics, AlbaNova, SE-106 91 Stockholm, Sweden}

\vspace{0.3 cm} Email: langmann@kth.se 

\end{center}


\begin{abstract}
I present the exact energy eigenstates and eigenvalues of a quantum
many-body system of bosons on non-commutative space and in a harmonic
oszillator confining potential at the selfdual point. I also argue
that this exactly solvable system is a prototype model which provides
a generalization of mean field theory taking into account non-trivial
correlations which are peculiar to boson systems in two space
dimensions and relevant in condensed matter physics. The prologue and
epilogue contain a few remarks to relate my main story to recent
developments in non-commutative quantum field theory and an addendum
to our previous work together with Szabo and Zarembo on this latter
subject.

\end{abstract}
\newcommand{\cG}{{\mathcal G}}
\newcommand{\cE}{{\mathcal E}}
\newcommand{\cF}{{\mathcal F}}
\newcommand{\cZ}{{\mathcal Z}}
\newcommand{\cH}{{\mathcal H}}
\newcommand{\cN}{{\mathcal N}}
\newcommand{\cD}{{\mathcal D}}
\newcommand{\nd}{{\phantom\dag}}
\newcommand{\cK}{{\mathcal K}}
\newcommand{\cC}{{\mathcal C}}

\renewcommand{\Re}{{\rm Re}}

\renewcommand{\Im}{{\rm Im}}

\newcommand{\X}{X} 
\newcommand{\x}{x}

\newcommand{\vk}{{\mathbf k}}

\newcommand{\Ref}[1]{(\ref{#1})}

\newcommand{\N}{{\mathbb N}}
\newcommand{\C}{{\mathbb C}}
\newcommand{\R}{{\mathbb R}}

\newcommand{\ii}{{\rm i}}
\newcommand{\ee}{{\rm e}}

\newcommand{\ndag}{{\phantom\dag}}

\begin{center}
\section*{Prologue}
\end{center}
{\it Recently the renormalizability of certain non-commutative (NC)
quantum field theory (QFT) models was established in important work by
Grosse and Wulkenhaar and the Orsay group
\cite{GW1,GW2,GW_review,O1,O2}; see Ref.\ \cite{O_review} for further
references and a lucid discussion of the significance of these
results. It seems that all examples of renormalizeable such models
share the duality property that their actions have the same form in
Fourier- as in position space \cite{LS}.  In this contribution I will
discuss the following models possessing this latter property
\cite{L1,L2,LSZ1,LSZ}:
\begin{equation}
 \left.\begin{array}{c} {\mathcal H} \\ S\end{array} \right\} =
 \int_{\R^{2n}} d^{2n}x\, \left( \Phi^\dag(x)[\sigma (-\ii\partial -
 B\cdot x)^2 +\tilde\sigma (-\ii\partial + B\cdot x)^2 -\mu ]\Phi(x)+
 \tilde g \Phi^\dag\star\Phi\star \Phi^\dag\star\Phi(x) \right). 
 \label{HS}
\end{equation} 
\noindent {\em [{\bf Notation:} I denote points in $\R^{2n}$,
$2n=2,4,\ldots$, by $x=(x^1,\cdots,x^{2n})$; $\Phi^{(\dag)}(x)$
represents a boson field to be specified in more detail below; I write
\begin{equation}
\label{Delta}
(-\ii\partial \pm B\cdot x)^2 \equiv \sum_{\mu=1}^{2n}\left(
-\ii\frac{\partial}{\partial x^\mu}\pm (B\cdot x)_\mu\right)^2 \;
\mbox{ with }\; (B\cdot x)_\mu \equiv \sum_{\nu=1}^{2n}B_{\mu\nu}x^\nu
\end{equation} 
and $B=(B_{\mu\nu})_{\mu,\nu=1}^{2n}$ some fixed skewsymmetric and
invertible $2n\times 2n$ matrix; $\sigma, \tilde\sigma\geq 0$, $\mu$
and $\tilde g$ are real parameters such that $\sigma+\tilde\sigma>0$;
$\star$ is the well-known Groenewold-Moyal product (see e.g.\
\cite{Szabo} for review) characterized by another skewsymmetric
$2n\times2n$ matrix $\theta=(\theta_{\mu\nu})_{\mu,\nu=1}^{2n}$ as
follows,
\begin{equation}
x^\mu\star x^\nu-x^\nu\star x^\mu = -2\ii \theta^{\mu\nu} ; 
\end{equation}
the dagger indicates complex conjugation and the Hilbert space
adjoint; $\N$ are the {\em positive} integers.]} I denote the same
mathematical expression by two different symbols since it can be
either interpreted as the Hamiltonian ($\cH$) of a quantum many-body
system on $2n$ dimensional space \cite{L1,L2} or, alternatively, as an
action ($S$) of a NC QFT model on $2n$ dimensional Euclidean spacetime
\cite{LS,LSZ1,LSZ}.  My discussion will be mainly restricted to the
special case where $B\theta=I$ (i.e.\ $\theta$ is the inverse of the
matrix $B$) where these models are exactly solvable \cite{L1,LSZ}.
One of my aims is to present results which have remained unpublished
up to now, another to add a few remarks to our previous publications
on this exactly solvable NC QFT model together with Szabo and Zarembo
\cite{LSZ1,LSZ} to point out an interesting alternative interpretation
of our results, and to emphasis that an interesting problem concerning
this model still remains to be solved. The new results are on the
exactly solvable quantum many-body system of bosons and presented in
sections \ref{s1.0}--\ref{s1.3}, and my remarks on the NC QFT model
are contained in the epilogue at the end.

Before going into my main story I shortly recall the simplification
arising at $B\theta=I$ \cite{L1} (in this discussion I will refer to
the mathematical object in \Ref{HS} as Hamiltonian $\cH$, but
everything I say applies word-by-word also to its interpretation as
action $S$). Obviously, the Hamiltonian in \Ref{HS} is the sum of two
term, $\cH=\cH_0+\cH_{int}$, where $\cH_0$ and $\cH_{int}$ are the
quadratic and quartic parts in the fields $\Phi^{(\dag)}$,
respectively. As is well-known, in many standard field theory models
one can expand the fields in a basis such that either the quadratic or
the quartic part of the Hamiltonian becomes simple, but in general it
is not possible to make both parts simple in the same basis, and this
is one main reason why, in general, field theory models are
computationally challenging (typically $\cH_0$ is simple in Fourier
space and $\cH_{int}$ in position space). However, for the model in
\Ref{HS} at the special point $B\theta=I$ there exists a basis in
which both, $\cH_0$ and $\cH_{int}$, are simple.  To be specific:
this latter basis is given by the common eigenfunctions $\phi_{\ell
m}(x)$ of the differential operators in \Ref{Delta} labeled by two
sets of integer vectors $\ell,m\in\N^n$, and by expanding the fields
in this basis
\begin{equation}
\label{expand}
\Phi(x)=\sum_{\ell,m} A^\ndag_{\ell m}\phi_{\ell m}^\dag(x),\quad
\Phi^\dag(x)=\sum_{\ell,m} A^\dag_{\ell m}\phi^\ndag_{\ell m}(x)
\end{equation} 
the Hamiltonian acquires the following remarkably simple form
\cite{L1,LSZ},
\begin{equation}
 \left.\begin{array}{c} {\mathcal H} \\ S\end{array} \right\} =
\sum_{\ell,m}(E_\ell+\tilde E_m-\mu)A_{\ell m}^\dag A_{\ell m}^\ndag +
g \sum_{\ell,m,\ell',m'} A^\dag_{\ell m'} A^\ndag_{\ell m}
A^\dag_{\ell' m} A^\ndag_{\ell' m'}.
 \label{HS1}
\end{equation} 
The parameters $E_\ell$ and $\tilde E_m$ are proportional to the
eigenvalues of the operators in \Ref{Delta} and given by
\begin{equation}
E_\ell= \sum_{j=1}^n 4 \sigma|B_j|(\ell_j-\frac12),\quad \tilde E_m =
\sum_{j=1}^n 4 \tilde \sigma|B_j|(m_j-\frac12)
\end{equation} 
with $\ell=(\ell_1,\ell_2,\ldots,\ell_n)$ and similarly for $m$; 
$|B_j|$ are eigenvalues of the matrix $\sqrt{B^2}$ (see \cite{LSZ} for
a precise statement), and
\begin{equation}
g = \frac{\tilde g}{\sqrt{\det(4\pi\theta)}}. 
\end{equation}
It is interesting to note that the model in \Ref{HS1} can be written
in the following matrix form,
\begin{equation}
 \left.\begin{array}{c} {\mathcal H} \\ S\end{array} \right\} = {\rm
Trace}( E A^\dag A + \tilde E A A^\dag -\mu A^\dag A + g (A^\dag A)^2)
 \label{HS2}
 \end{equation}
where $A$, $E$ and $\tilde E$ above stands for the infinite matrices
with matrix elements $A_{\ell m}$, $E_\ell\delta_{\ell m}$, and
$\tilde E_m\delta_{\ell m}$, respectively, the matrix adjungation is
defined such that $(A^\dag)^\ndag_{\ell m}=A^\dag_{m\ell}$, and matrix
multiplication is understood.  }

\setcounter{section}{0} 
\section{Introduction}
\label{s1.0}
Interacting boson systems have been of interest in theoretical physics
since the early days of quantum physics, and a recent increased
interest in this subject was triggered by remarkable experimental
progress to realize and study the Bose-Einstein condensation; see
e.g.\ \cite{PS} for a recent text book in this topic. I believe that
these developments provide a good additional motivation for studying
the NC quantum many-body Hamiltonian $\cH$ in \Ref{HS1}: as I will
argue in more detail in my first remark in section~\ref{s1.3}, this
Hamiltonian defines a prototype model which allows to study a
particular type of correlations and its effect on the Bose-Einstein
condensation in an exact solution.

In the main part of this paper I thus interpret the model in \Ref{HS}
as Hamiltonian $\cH$ of bosons moving on $2n$ dimensional space
$\R^{2n}$ and interacting with a particular four point interaction. I
will show that this model is exactly solvable in the sense that all
its energy eigenstates and eigenvalues can be computed explicitly. As
will be seen, this exact solution provides an example of a correlated
boson system. To simplify notation and to allow for a simple physical
interpretation I restrict my discussion to the case $2n=2$ and
$\sigma=\tilde\sigma=1/4$, but my results can be straightforwardly
generalized to $2n>2$ and general parameter values. The parameter
$\mu$ corresponds to the chemical potential, and, for my purposes, one
can assign to it any convenient value.

I mention in passing that the fermion variant of this model was
introduced and analyzed in \cite{L1,L2} but, to my knowledge, the
boson story presented here has not appeared in the literature before.

The plan of the rest of this paper is as follows. In
section~\ref{s1.1} I give a precise definition of the quantum
many-body model, and in section~\ref{s1.2} I present its
solution. Section~\ref{s1.3} contains various remarks, and, in
particular, I explain there why I believe that this model is a
prototype model for interacting bosons and relevant in condensed
matter physics.

\section{Definition of the model}
\label{s1.1}
I consider the quantum many-body system defined by the Hamiltonian in
\Ref{HS} where the boson fields $\Phi^{(\dag)}(x)$ are operators
acting on a boson Fock space $\cF$ defined by the usual canonical
commutator relations and a normalized vacuum state $\Omega$
annihilated by all operators $\Phi(x)$; see e.g.\ \cite{NO}.
Expanding the fields as in \Ref{expand} these latter relations are
equivalent to
 \begin{equation}
 [A^{\ndag}_{\ell m},A^{\dag}_{\ell' m'}] = \delta^{\ndag}_{\ell
 \ell'}\delta^{\ndag}_{mm'},\quad [A^{\ndag}_{\ell m},A^{\ndag}_{\ell'
 m'}] =0,\quad A^{\ndag}_{\ell m}\Omega=0
 \end{equation}
for all $\ell,m,\ell',m'$, as usual. Choosing $B\theta=I$ 
this Hamiltonian can be written as
\begin{equation}
\label{cH} 
 \cH = \cH_0+\cH_{int},\quad \cH_0= \sum_{\ell,m}(E_\ell + \tilde
E_m) A_{\ell m}^\dag A_{\ell m}^\ndag,\quad \cH_{int} = g
\sum_{\ell,m,\ell',m'} A^\dag_{\ell m'} A^\dag_{\ell' m} A^\ndag_{\ell
m} A^\ndag_{\ell' m'}
 \end{equation}
where I found it convenient to normal order the interaction term $
\cH_{int}$ (this corresponds to a renormalization of $\mu$ which can
be ignored) and rename $E_\ell-\mu$ to $E_\ell$. To simplify my
discussion I set $2n=2$, $\sigma=\tilde\sigma=1/4$ and $\mu= |B|$ so
that $\ell,m\in\N$ and
\begin{equation}
\label{ek}
E_\ell + \tilde E_m = |B|(\ell+m-2).
\end{equation}
Then the model describes interacting bosons on two dimensional space
and confined by harmonic oszillator potential, with $\ell-1$ and $m-1$
the usual harmonic oscillator quantum numbers. A useful alternative
interpretation of the quantum numbers $\ell$ and $m$ is as $x$- and
$y$ components of (quasi-)momenta of bosons in two dimensions. In this
latter interpretation one is interested in other dispersion relations
like $E_\ell + \tilde E_m\propto (\ell^2 + m^2)$ and $\ell,m$ running
also over negative integers (see the first remark in
section~\ref{s1.3}), and it is therefore important to note that such
changes do not affect the exact solubility of the model. Then the
interaction term describes two body scattering processes where two
bosons with initial momenta $(\ell, m)$ and $(\ell ', m')$ exchange
the $y$-components of their momenta while the $x$-components remain
the same, or vice versa \cite{L2}.

\section{Exact solution}
\label{s1.2}
I now discuss how to construct exact energy eigenstates of this
model. For that I consider the quasi-free states (the
normalization of the eigenstates will be ignored)
\begin{equation}
\label{eta}
\eta = A^\dag_{\ell_1 m_1}A^\dag_{\ell_2 m_2}\cdots A^\dag_{\ell_N
m_N}\Omega
\end{equation} 
for fixed quantum numbers $\ell_j$ and $m_j$ in $\N$, with $N$ an
arbitrary fixed non-negative integer.  One can interpret this as a
state containing $N$ bosons with momenta $(\ell_j,m_j)$.  Each such
state is an eigenstate of the quadratic part $\cH_0$ of the
Hamiltonian, and the corresponding eigenvalue is
\begin{equation}
\cE_0 = \sum_{j=1}^N (E_{\ell_j} + \tilde E_{m_j} ).
\end{equation}
I will refer to this as kinetic energy. 

It is important to note that the permutation group $S_N$ of $N$
elements has a natural action on these states $\eta$ as follows,
\begin{equation}
\label{Peta}
P\eta :=  A^\dag_{\ell_1 m_{P1}}A^\dag_{\ell_2 m_{P2}}\cdots
A^\dag_{\ell_N m_{PN}}\Omega
\end{equation}
for all $P\in S_N$, and that all these states $P\eta$ are degenerate
eigenstates of $\cH_0$. Moreover, one can show that the action of the
interaction part of the Hamiltonian $\cH_{int}$ on such a state $\eta$
is
\begin{equation}
\label{Hint}
\cH_{int}\eta = 2 g \sum_{1\leq j<k\leq N} T_{jk} \eta
\end{equation}
where $T_{jk}\in S_N$ is the transposition which interchanges $j$ and
$k$ and leaves all other integers $1,2,\ldots, N$ the same. One can
interpret $T_{jk}$ as the operator exchanging the $y$-components of
the momenta of the $j$-th and the $k$-th boson leaving the
$x$-components the same, or vice versa. Obviously this implies that
all eigenstates of $\cH$ are of the form
\begin{equation}
\Psi = \sum_{P\in S_N} a_P P\eta
\end{equation}
for some $\eta$ and certain coefficients $a_P$ to be determined. I 
will get back to the problem of how to construct all these eigenstates
and corresponding eigenvalues further below. 

For now I consider particular such eigenstates which can be obtained
by elementary methods and which include the groundstates in the weak-
and strong coupling limits. These eigenstates are given by
\begin{equation}
\label{etapm}
\eta^\pm = \sum_{P\in S_N} (\pm )^{P} P \eta
\end{equation} 
where $(+)^{P}$ is always $1$ and $(-)^{P}=1$ for even and $-1$ for
odd permutations $P$, respectively. To see that these are eigenstates
we note that $T_{jk}\Psi_\pm=\pm\Psi_\pm$, which implies $\Psi^\pm$ is
an exact eigenstate of $\cH_{int}$ with eigenvalue $\pm g N(N-1)$
(since $\sum_{1\leq j<k\leq N}=N(N-1)/2$), and thus
\begin{equation}
\cH\eta^\pm = (\cE_0 \pm g N(N-1))\eta^\pm.
\end{equation} 
It is interesting to note that the state $\eta^-$ has a fermion-like
character and, as discussed below, this implies a strong variant of
the Pauli exclusion principle which will play an important role.  As
will be shown further below, the states $\eta^\pm$ are extremal in the
sense that they have the largest possible interaction energies.

In particular, for $g \leq 0$ the groundstate of the model at fixed
particle number $N$ is the state $\eta^+$ such that the kinetic energy
$\cE_0$ assumes its smallest possible value. It is easy to see that
the state in \Ref{eta} with the minimum kinetic energy is
\begin{equation}
\label{eta0}
\eta_1 = (A^\dag_{1,1})^N\Omega ,  
\end{equation} 
i.e.\ all bosons are in the same one-particle state with momentum
$(\ell,m)=(1,1)$.  Note that $\eta_1$ is the well-known Bose-Einstein
condensate (BEC) groundstate of the non-interacting system ($g=0$). In
fact, this state is the groundstate for all $g<0$ (this is true since
$\eta_1^+$ equals $\eta_1$ up to a constant), and it is easy to see
that the corresponding groundstate energy is
\begin{equation}
\label{cE1} 
\cE_1 = g N(N-1). 
\end{equation}

For $g>0$ the states $\eta^+$ have a large interaction energy, and for
sufficiently large $g>0$ the groundstate of the model should be the
state $\eta^-$ with $\eta$ such that the kinetic energy $\cE_0$ is
minimal. It is important to note that one now cannot take as $\eta$
the state $\eta_1$ in \Ref{eta0} since $\eta_1^-$ vanishes. More
generally, the following strong variant of the Pauli exclusion
principle holds true: {\it The state $\eta^-$ in \Ref{etapm} is
non-zero only if all the $x$- and all the $y$-components $\ell_j$ and
$m_j$ of the boson momenta in the state $\eta$ in \Ref{eta} are
distinct.}\footnote{The standard Pauli principles for fermions is
weaker since it only requires that all the momenta $(\ell_j,m_j)$ are
distinct.} [{\em Proof:} Consider a state $\eta$ in \Ref{eta} such
that $\ell_j=\ell_k$ and/or $m_j=m_k$ for some $j<k$. This implies
$T_{jk}\eta=\eta$, but then $\eta^-=\sum_P(-)^P P\eta = \sum_{P}(-)^P
P T_{jk}\eta=-\eta^-$, and thus $\eta^-=0$.]  A state whose momenta
are all distinct and which has the lowest possible kinetic energy is
\begin{equation}
\label{eta1} 
\eta_2 = A^\dag_{1,1}A^\dag_{2,2}\cdots A^\dag_{N,N}\Omega, 
\end{equation}
and thus 
\begin{equation}
\label{etam1} 
\eta_2^- = \sum_{P\in S_N}(-1)^P 
A^\dag_{1,P1}A^\dag_{2,P2}\cdots A^\dag_{N,PN}\Omega
\end{equation}
is the groundstate of the model in the strong coupling limit. The
corresponding groundstate energy is
\begin{equation}
\label{cE2} 
\cE_2  = |B|N(N-1) - g N(N-1) 
\end{equation}
(since $\sum_{\ell=1}^N (\ell-1) = N(N-1)/2$).  As discussed below,
$\eta_2^-$ is actually the groundstate of the model not only in the
strong coupling limit but for all $g\geq |B|$.

I now discuss the problem of finding the groundstate for intermediate
coupling values. Note that, for fixed $\eta$ in \Ref{eta}, the states
$P\eta$ in \Ref{Peta}, $P\in S_N$, span a subspace $\cF_{\eta}$ of the
boson Fock space $\cF$. The dimension of this subspace is $\leq N!$,
and it is $N!$ if and only if all the $x$- and $y$-components $\ell_j$
and $m_j$ of the bosons in the state $\eta$ are distinct. It is
important to note that \Ref{Peta} defines a representation of the
permutation group $S_N$ on $\cF_\eta$, and this representation is, in
general, reducible. Moreover, the operator
$$
C_N = \sum_{1\leq j<k\leq N} T_{jk}
$$ appearing in \Ref{Hint} commutes with all permutations $P\in S_N$,
and it is therefore a constant in each irreducible representation
(irrep) of $S_N$. Since on $\cF_\eta$ the kinetic energy $\cH_0$ is
constant and the interaction $\cH_{int}$ proportional to $C_N$, {\it
the problem of constructing eigenstates of $\cH$ is equivalent to
decomposing the representation of $S_N$ on $\cF_\eta$ described above
into irreps.}  This is a classical problem solved in group theory; see
e.g.\ Chapter IV in \cite{Weyl}: the irreps of $S_N$ can be labeled by
partitions $\lambda$ of $N$, i.e.\
$\lambda=(\lambda_1,\lambda_2,\ldots,\lambda_K)$ with integers
$\lambda_j$ such that
\begin{equation}
\lambda_1\geq \lambda_2\geq \cdots \geq \lambda_K>0,\quad
\sum_{j=1}^K\lambda_j = N,
\end{equation}
and the value of $C_N$ in an irrep $\lambda$ is
$C_N=\sum_{j=1}^K[\frac12\lambda_j(\lambda_j+1) -j\lambda_j]$; see
e.g.\ Eq.\ (4-3) in \cite{Chen}. Moreover, the states in an irrep
$\lambda$ can be obtained by applying to states $\eta$ the so-called
Young symmetrizer \cite{Weyl} denoted by $Y^{\lambda}$. One thus 
concludes that the energy eigenstates of this model are
$Y^{\lambda}\eta$ with the corresponding eigenvalues
\begin{equation}
\cE = \sum_{j=1}^N  (E_{\ell_j}+\tilde E_{m_j}) + \sum_{j=1}^K
g\lambda_j(\lambda_j+1-2j). 
\end{equation}
The states $\eta^+$ and $\eta^-$ in \Ref{etapm} correspond to the
special cases $\lambda=(N)$ and $\lambda=(1,1,\ldots,1)\equiv (1^N)$,
respectively. In principle this gives all eigenstates and eigenvalues
of the model. There is, however, an important complication: as seen in
the previous section for the special case $\lambda=(1^N)$, if there
are degeneracies many of the eigenstates $Y^{\lambda}\eta$ vanish. To
find the groundstate of the model we therefore must determine the
state $\eta$ in \Ref{eta} of minimal kinetic energy and such that, for
a fixed partition $\lambda$, $Y^{\lambda}\eta$ is non-zero. This
problem has the following solution,
\begin{equation}
\label{c1}
\eta = 
(A_{1,1}^\dag)^{\lambda_1} (A_{2,2}^\dag)^{\lambda_2}\cdots
(A_{K,K}^\dag)^{\lambda_K}\Omega, 
\end{equation}
and the smallest possible energy eigenvalue in an irreps $\lambda$ is
therefore
\begin{equation}
\label{cE}
\cE_\lambda = \sum_{j=1}^K \lambda_j\left(E_j + \tilde E_j +
g(\lambda_j+1 - 2 j) \right) .
\end{equation}
One can determine the groundstate of the model by finding the
partition $\lambda$ of $N$ which minimizes the energy in \Ref{cE}.
The solution of this problem depends on the dispersion relation
$E_\ell+\tilde E_m$. Using the one in \Ref{ek} one finds $\lambda=(N)$
for $g\leq 0$ and $\lambda=(1^N)$ for $g\geq |B|$, which confirms that
the states in \Ref{eta0} and \Ref{eta1} are the groundstates for
$g\leq 0$ and $g\geq |B|$, respectively. In the intermediate regime
$0<g<|B|$ the groundstate is given by a partition approximated by
\begin{equation}
\label{partition}
\lambda_j \simeq \alpha(K+1 - j),\quad K \simeq
\sqrt{\frac{2N}{\alpha}} \mbox{ with } \alpha \simeq \frac{|B|}g-1 >
0
\end{equation}
where "$\simeq$" means that the l.h.s.\ is the non-negative integer
closest to the r.h.s., and this approximation becomes exact in the
limit when the boson number $N$ becomes infinite.

I finally note that one can prove that the eigenstates of this model
are, in general, correlated by finding one non-zero connected 4-point
correlation function. As an example I consider the normalized strong
coupling groundstate for $N=2$:
\begin{equation}
\Psi = \frac1{\sqrt2}\left(A_{1,1}^\dag A_{2,2}^\dag - A_{1,2}^\dag
A_{2,1}^\dag\right)\Omega, 
\end{equation}
which supports the following non-trivial connected 4-point correlation
function,
\begin{equation}
\label{example}
(\Psi,A^\dag_{1,2} A^\dag_{2,1}A^\ndag_{1,1}
  A^\ndag_{2,2}\Psi)-(\Psi,A^\dag_{1,2}A^\ndag_{1,1}\Psi)(\Psi
  A^\dag_{2,1}A^\ndag_{2,2}\Psi)
  -(\Psi,A^\dag_{1,2}A^\ndag_{2,2}\Psi)(\Psi
  A^\dag_{2,1}A^\ndag_{1,1}\Psi) = -\frac12
\end{equation}
with $(\cdot,\cdot)$ the inner product in the boson Fock space.

\section{Concluding remarks}
\label{s1.3}
\begin{enumerate}
\item A key problem in theoretical physics is to do reliable
computations in quantum models with interactions so large that
perturbation theory does not apply. One well-known and often
successful strategy in this context is mean field theory. It is
interesting to note that one approach to mean field theory is to
truncate the interaction in the model under consideration and only
keep the so-called Hartree- and Fock terms, which typically leads to an
exactly soluble model whose solution is equivalent to mean field
theory of the original model; see \cite{L2} for a discussion of this
in the context of interacting fermion systems.  Mean field theory does
not take into account correlations, and the latter are believed to be
particularly important in two spatial dimensions (2D). {\it I propose
that the Hamiltonian in \Ref{cH} defines a prototype model allowing to
study important 2D correlations in an exact solution.}  To motivate
this I consider the following standard 2D boson model
\begin{equation}
\label{model}
\cH = \sum_{\vk} \frac{\vk^2}{2M} b^\dag(\vk) b(\vk) + \frac{U}{L^2}
\sum_{\vk_1,\vk_2,\vk_3,\vk_4}\delta_{\vk_1+\vk_2,\vk_3+\vk_4}
b^\dag(\vk_1)b^\dag(\vk_2)b(\vk_3) b(\vk_4)
\end{equation}
with the boson mass $M>0$ and coupling parameter $U>0$. The boson
operators $b^{(\dag)}(\vk)$ are labeled by 2D momenta
\begin{equation} 
\vk=(k_x,k_y),\quad k_x,k_y\in \frac{2\pi}{L}{\mathbb Z} \; \mbox{
such that } \; |k_{x,y}|<\frac{\pi}a
\end{equation} 
and obey the usual relations,
$[b(\vk),b^\dag(\vk')]=\delta_{\vk,\vk'}$ etc. The parameters $L \gg
a>0$ correspond to the system size ($L$) and a lattice constant ($a$)
and provide a IR and UV cutoff for the model.  The interaction term in
this Hamiltonian comes from a local interaction in position space and
describes scattering processes where two bosons with momenta $\vk_3$
and $\vk_4$ are scattered into states with momenta $\vk_1$ and
$\vk_2$, and the model is complicated since all possible such
scattering processes occur with equal strength and are restricted only
by overall momentum conservation.  The Hartree- and Fock terms
correspond to the scattering terms where $\vk_1=\vk_4$, $\vk_2=\vk_3$
and $\vk_1=\vk_3$, $\vk_2=\vk_3$, and they are (essentially) trivial
for this model in the sense that they only add an energy $\simeq
4gN^2$ and do not (much) affect the groundstate. Note that the
interaction contains also the scattering terms where
\begin{equation}
(k_1)_x=(k_4)_x,\quad (k_2)_x=(k_3)_x,\quad (k_1)_y=(k_3)_y,\quad
(k_2)_y=(k_4)_y
\end{equation}
and similar terms with $x$ and $y$ interchanged.  These scattering
terms are Hartree-like in the $x$- and Fock-like in the $y$-component
of the momenta and vice versa, and they are peculiar to 2D.  {\em If
one restricts the interaction terms in the Hamiltonian in \Ref{model}
and only includes these latter mixed Hartree-Fock terms one obtains
exactly a Hamiltonian as in \Ref{cH} with 
\begin{equation}
E_\ell+\tilde E_m =\frac1{2M}\left(\frac{2\pi}L \right)^2 (\ell^2 +
m^2),\quad g=\frac{2U}{L^2}
\end{equation} 
and integers $\ell,m$ such that $|\ell|,|m|<L/(2a)$.} As mentioned,
this latter truncation is very similar to a successful method to
derive useful mean field theories for interacting fermion systems, and
I thus regard the model in \Ref{cH} as a generalized mean field
model. The exact solution of this model above does not rely on the
form of $E_\ell+\tilde E_m$ (except for the groundstate, of course),
and, as argued below, this model describes interesting ``physics''
which cannot be accounted for in mean field theory.
\item It is interesting to note how the character of the groundstate
of the model changes with increasing coupling constant $g$: for $g=0$
one has the standard BEC groundstate in \Ref{eta0} where all bosons
are in the same one particle state $(\ell_j,m_j)=(1,1)$. As the
coupling increases it becomes more favorable to reduce the
degeneracies and thus the number of bosons in the BEC, and one finds a
distribution of the momenta as described by the partition in
\Ref{partition} and a correlated groundstate. Moreover, the BEC
condensate in the ground state for $2|B|/(N+2)<g<|B|$ is
\begin{equation}
\langle A_{1,1}^\dag A^\ndag_{1,1}\rangle \simeq K\alpha \simeq
\sqrt{2N\left(\frac{|B|}g-1\right)},
\end{equation}
and it becomes 1 for $g\geq |B|$ where the groundstate becomes
maximally correlated. Moreover, as demonstrated in \Ref{example} above
in a simple example, one can construct non-trivial 4-point Green's
functions for the model to prove that its groundstate is, in general,
correlated and thus not accessible by mean field theory.  I hope that
these remarks are sufficient to convince the reader that the
``physics'' of this model is non-trivial and interesting.  I plan to
present a more detailed discussion elsewhere.
\item It is important to note that the model in \Ref{cH} describes a
stable system only in the parameter regime $0\leq g\leq |B|$ since
otherwise the groundstate energy can be decreased by increasing the
particle number $N$ to infinity (this follows from \Ref{cE1} and
\Ref{cE2}). In my interpretation of this model as generalized
Hartree-Fock model the instability for $g>|B|$ is removed by the
Hartree-Fock energy $\simeq 4gN^2$ which should also be included.
\item Obviously nearly everything I wrote in the previous section can
be immediately generalized to $2n>2$ and other values for $\sigma$ and
$\tilde\sigma$, and the only change will be the solution to the
problem to minimize the energy in \Ref{cE}.
\item In this paper I only computed the groundstate of the model and
demonstrated how to compute the other energy eigenstates and
eigenvalues. Obviously it would be interesting to compute also other
quantities, like Green's functions and the partition function.
\end{enumerate} 
 
\begin{center}
\section*{Epilog}
\end{center}
{\it The two models in \Ref{HS} are closely related: the NC QFT model
defined by $S$ can be obtained as infinite temperature limit of the
quantum many-body model $\cH$. Indeed, one can write the generating
function for the Green's functions of the latter model as matrix path
integral (where the integration variables $A^{(\dag)}_{\ell m}(\tau)$
are periodic functions of the Matsubara time $\tau\in[0,\beta]$ with
$\beta$ the inverse temperature; see e.g.\ \cite{NO}), and the
functional integral defining the NC QFT model \cite{LSZ} can be
obtained as a limit $\beta\to 0$ from that. Thus the model $\cH$ in
\Ref{HS} defines a $2n+1$ dimensional QFT.  It would be interesting to
use this relation to defer from my results on the latter model results
for the former model.

The NC QFT model $S$ in \Ref{HS} has interesting and non-trivial QFT
divergences which one has to treat by regularization and
renormalization. I believe that this models provides an interesting
example where the role of such divergences can be studied in detail
and by exact and explicit results beyond perturbation theory, and that
our previous results on this \cite{LSZ1,LSZ} are only a first step in
this direction: as I will argue below, there are other QFT limits than
the ones studied in this latter work, and one of these other limits is
more interesting and more difficult than the others. The explicit
solution of the model in this latter limit is a challenging but doable
project for the future.

To be more specific: The natural regularization for the NC QFT model
$S$ in \Ref{HS1} is to restrict the fields $A_{\ell m}^{(\dag)}$ to
$\ell=(\ell_1,\ell_2,\ldots,\ell_n)$ such that
\begin{equation}
\ell_j=1,2,\ldots, L<\infty  
\end{equation}
and similarly for $m$. Then the fields $A_{\ell m}$ can be naturally
interpreted as components of a \footnote{Note that the symbol $N$ in
the following and in sections~\ref{s1.0}-\ref{s1.3} have different
meanings!}  $N\times N$ matrix $A$ with $N=L^n$.  With that the
functional integral defining the NC QFT model becomes a well-defined
integral over $\R^{2N^2}$, and the non-trivial task is to find a
dependence of the model parameters $\sigma$, $\tilde\sigma$, $-\mu$,
and $g$ on the cut-off parameter $N$ such that the limit $N\to\infty$
is well-defined an non-trivial.

In \cite{LSZ} we studied two such limits for the case $\tilde\sigma=0$
which we called IR- and UV limit: The IR limit corresponds to the
following scaling of parameters,
\begin{equation}
\label{IRlimit}
\sigma=1,\quad g=\frac{g_{ren}}N,\quad B = \frac{B_{ren}}{N^{1/n}}
,\quad \mu=\mu_{ren}
\end{equation} 
where the parameters with the subscript ``{\it ren}'' (short for
``renormalized'') are independent of $N$. The results for the Green's
functions in this limit can be found in \cite{LSZ}. I only mentioned
here that $B\to 0$ for $N\to \infty$ leads to a 2-point Green's
function which is translational invariant, $G(x;y)=G(x-y;0)$, and all
the higher Green's functions are trivial. In the IR limit the duality
symmetry of this model \cite{LS} is broken, which implies the
existence of a dual limit where $B\to \infty$ as $N\to\infty$ and with
Green's function obtained from the ones in the IR limit by a duality
transformation. The 2-point Green's function in this latter UV limit
is non-trivial and ultra-local, $G(x;y)\propto \delta^{2n}(x-y)$, and
the higher Green's functions are again trivial \cite{LSZ}.

Now comes my addendum to \cite{LSZ}: {\em It is possible to get a
third limit in which the above mentioned duality symmetry is not
broken as follows: rather then keeping $\sigma$ constant and scaling
$B$ like $N^{-1/n}$ one can scale $\sigma$ like $N^{-1/n}$ and keep
$B$ constant in the limit $N\to\infty$:
\begin{equation}
\sigma=\frac{\sigma_{ren}}{N^{1/n}},\quad g=\frac{g_{ren}}N,\quad B
=B_{ren},\quad \mu= \mu_{ren}.
\end{equation} 
The non-trivial scaling of $\sigma$ can be interpreted as
multiplicative regularization.  It is easy to deduce from the results
in \cite{LSZ} the 2-point Green's function in this third limit,
\begin{equation}
G(x;y) = \sum_{\ell,m} \langle A^\dag_{\ell m} A^\ndag_{\ell m}\rangle
\phi^\ndag_{\ell m}(x)\phi^\dag_{\ell m}(y)
\end{equation}
with $\langle A^\dag_{\ell m} A^\ndag_{\ell m}\rangle$ depending only
on $\ell$ and computed explicitly in \cite{LSZ}, section 4.2; the
higher Green's functions are again trivial.}

I finally would like to emphasis that the limits described above are
restricted to the case $\tilde\sigma=0$, and for $\tilde\sigma>0$
there should exist another limit leading to non-trivial higher Green's
function and which should describe a non-trivial fixed point of the
renormalization group of the models in \Ref{HS}. I expect that this
latter limit is the one studied in a closely related model in
\cite{GW4,O4} (this latter model is similar to ours for
$\sigma=\tilde\sigma$). It is certainly not easy but, as I believe,
possible and very desirable to compute the Green's functions of the
model in this QFT limit explicitly.  In this context I should mention
the non-perturbative renormalization of NC $\phi^3$-theory which was
recently established by Grosse and Steinacker \cite{GS}.}

\bigskip

\noindent {\bf Note added:} The model $\cH$ in \Ref{HS} for $2n=2$,
$\sigma=\tilde\sigma=1/2$, and at finite temperature was recently
studied in \cite{FS}.

\bigskip

\subsection*{Acknowledgments}
I would like to thank Jonas de Woul for helpful discussions and for
reading the manuscript. This work was supported by the Swedish Science
Research Council (VR) and the European Union through the FP6 Marie
Curie RTN {\em ENIGMA} (Contract number MRTN-CT-2004-5652).


\end{document}